# A risk assessment of downdrag induced by reconsolidation of clays after upwards pipe jacking

N.A. Labanda & A.O. Sfriso
*SRK Consulting, Argentina*
*Universidad de Buenos Aires, Argentina*

D. Tsingas & R. Aradas
*Jacobs, Argentina*

M. Martini
*Salini-Impregilo S.p.A., Italy*

ABSTRACT: Salini-Impregilo is building part of the largest sanitary sewer system in the history of Argentina in the suburbs of Buenos Aires City, to serve a population of almost five million people. The project is an outfall TBM tunnel 12 km long, starting from a reception shaft in the river margin, and transporting the sewage 35 meters below the *Rio de la Plata* riverbed to the point of discharge. Within the final kilometer of the tunnel, a set of 36 standing pipes so-called *risers* are constructed by driving steel tubes upwards and passing through dense sands, sandy clays and soft clays. Risers are linked-up with the launching lining segment using flange unions.
Driving of risers upwards will generate excess pore pressure and disturbance in fine soils and, once the pipe is placed in its final position, negative skin friction due to reconsolidation and creep. A risk assessment of the downdrag is presented in this paper, based on the estimation of the force and/or displacement in the riser-tunnel union generated by this effect. The issues of whether it is desirable to instalock the riser-tunnel union at an early age after installation of the riser and the time lapse required to reduce negative skin friction effects are discussed. Results are validated by comparing the model results with field measurements in prototype models.

## 1 INTRODUCTION

The *Matanza-Riachuelo river* flows along the riparian lands between the City and the Province of Buenos Aires (Argentina). It is a water stream 64 km long and 35 m wide in average that flows into the *Rio de la Plata*. The Riachuelo river is considered as one of the most polluted water courses in the world, ruined across the years by the discharge of several kind of untreated industrial and waste water.

The *Matanza-Riachuelo basin recovery project* is an ambitious plan to manage the wastewater from the left margin of the *Rio de la Plata* through a sewerage interceptor tunnel, a treatment plant and a discharge tunnel into the *Rio de la Plata*. The project is divided into several contracts. One of them, named *Lot 3*, holds the outfall tunnel, an EPB-TBM tunnel running 35 meters below the riverbed of the *Rio de la Plata*. A scheme of the plant together with the design of the system at conceptual level is shown in Figure 1.

The tunnel has an internal diameter of 4.3 m and a total length of 12 km including a 1.5 km diffusor zone with of 34, 28 m long standpipes named '*risers*' that daylight in the riverbed. Risers are lifted from the interior of the tunnel by driving upwards steel tubes. A construction sequence of risers installation is shown in Figure 2. Driving the risers will generate excess pore pressure and disturbance in the fine surrounding soils and, once the pipe is placed in its final position, negative skin friction due to reconsolidation and creep of soft clay layers. Due to an optimization of the design, the tunnel elevation was lifted and, consequently, the length of the risers reduced from 30.5 m to 28.0 m, thus reducing the thickness of sandy layers to be crossed. While this optimization is beneficial to reduce the jacking forces through dense sands, it does worsen the downdrag phenomenon due to the loss of the positive skin friction provided by the sands. This could generate two scenarios:

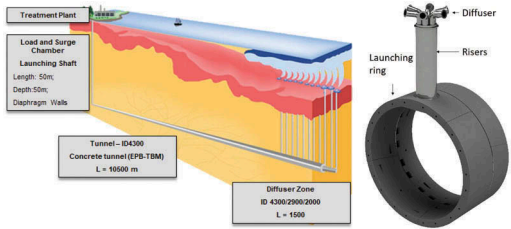

Figure 1. Conceptual design of the Matanza-Riachuelo outfall. Detail of launching rings and risers in diffusor zone.

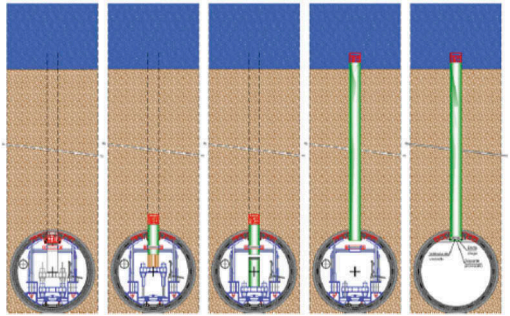

Figure 2. Risers construction stages.

- Scenario N°1: If the riser *is not* bolted to the tunnel, a vertical displacement is experienced by the pipe, slightly increasing the local hydraulic loss.
- Scenario N°2: If the riser *is* bolted to the tunnel, the negative skin friction will produce a non-negligible force acting on the crown lining segment.

A risk assessment using a simplified model to estimate displacements (scenario N°1) or forces (scenario N°2) produced by negative skin friction in risers due to consolidation and creep is presented in this paper. A brief introduction to the geotechnical site characterization is presented in Section 2, while the constitutive models employed are presented in 3. The numerical model and its results about downdrag, structural forces in risers and comparisons with field test results are shown in 4. Finally, some conclusions are drawn in Section 5.

## 2 GEOTECHNICAL SITE CHARACTERIZATION

The mentioned project is placed below the *Rio de la Plata*, where first 20 to 25 meters depth from the riverbed are composed by normally consolidated soft clays and silty clays stemmed by fluvial depositions, known as the '*Post-pampeano*' formation. This formation is characterized by a liquid limit $\omega_L$ varying from $\sim 30$ to $\sim 60$, plasticity index from $\sim 10$ to $\sim 30$, $\frac{S_u}{\sigma'_v} \sim 0.24|0.28$ and fine content above the 85% (Sfriso 1997).

The '*Puelchense*' formation can be found below the post-pampeano, between 27 to 31 meters depth from the riverbed, composed by a dense clean sands with an in-situ relative density above 80%, an effective friction angle in constant volume $\phi_{cv} \sim 32°$ and a dilatancy angle $\psi \sim 6°|8°$.

Between the puelchense and post-pampeano formation, a transition of 5 m to 11 m thickness composed by interleaved layers of soft clays and medium to dense silty sands is encountered. Figure 3 shows a detailed stratigraphy of the site, where the transition layer is represented as fundamentals soil layers classified by CPTu in-situ test and Robertson charts (Roberton 2009). It also shows the trace of the mentioned tunnel, the position of the CPTu tests used to calibrate the numerical models and, in red, the considered risers for our analysis. In order to illustrate the soil classification in the transition, Figure 4 plots the Robertson chart for the layer in CPTu-01A and CPTu-05A. It can be seen that the mentioned stratum is mainly composed by normally and under consolidated soils, with an erratic distribution of soft clays and sandy soils. The underconsolidation is produced by a constant upflow coming from the puelchense formation, a confined aquifer with an over pressure with respect to the river of, approximately, 2 m. It is clear in the site profile that, as the crown elevation is lifted, the thickness of sandy soils across the risers decreases. This is crucial for our problem because, as is demonstrated in this paper, sandy soils provides a confinement pressure in the pipe diminishing displacement or forces in the riser-tunnel union.

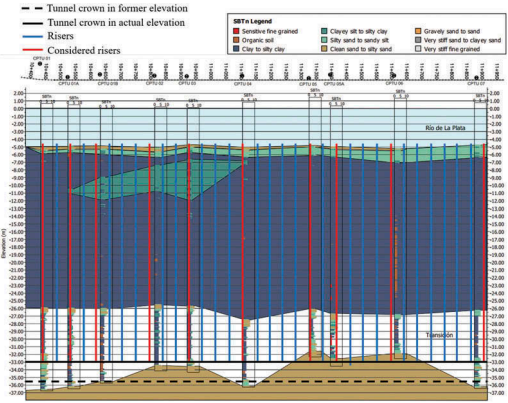

Figure 3. Site stratigraphy, risers location and crown elevation.

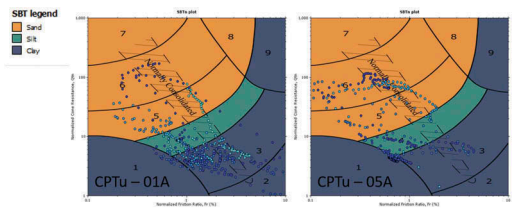

Figure 4. Robertson charts for the transition for CPTu-01A and CPTu-05A.

## 3 CONSTITUTIVE MODELLING

Soft clays in the riverbed of ther *Rio de la Plata* are anisotropically normally consolidated soils, sufficiently ancient to assume that primary and secondary consolidation has been occurred. Nevertheless, during risers driving, an intensive soil remolding is carried out in the surroundings of the pipe, till a distance of 4 to 5 times its diameter. The soil disturbance, generated at constant humidity, produce excess pore pressure erasing the stress history of the material. Once the driving is ended, excess pore pressure dissipation and reconsolidation in the influenced area, produce downdrag forces and, consequently, negative friction in the riser. Soft soil creep model is used for clays to take into account the time-dependent behavior produced by remolding. At the same time sandy layers lose densification, approaching the effective friction angle to the constant volume friction angle, reducing the dilatancy angle to $0°$ and, consequently, its confinement potential. Hardening soil small is used to simulate the mechanical behavior of this kind of soils.

Figure 5 shows experimental results of compressibility index as a function of the liquid limit, obtained by unidimensional consolidation test for the post-pampeano formation, with a comparison with Terzaghi and Skempton correlations (Sfriso 1997). The swelling index is considered as $C_s \approx \frac{1}{5} | \frac{1}{10} C_c$ and the secondary compressibility index for the mentioned formation is $C_\alpha = 0.013$ (Ledesma 2008). Parameters for sandy soils are obtained from classical correlations based on the standard penetration test (SPT) and the transition layer is simulated as a composition of soft clays and sands layers. The geotechnical profile for each riser is calibrated by comparing pore pressures calculated with the numerical model and CPTu measurements. Table 1 summarize constitutive models and parameters used for our simulations. Figure 6 shows a comparisons between consolidated-drained triaxial test of samples taken from the post-pampeano formation and the constitutive model calibration. Soft soil creep has a limitation to reproduce the mechanical behavior in low pressures in low strains. Nonetheless in this particular problem, the numerical model works in large strains and the proposed constitutive model reproduce reasonably the maximum deviatoric resistance of soft clays. Figure 6 shows a comparisons between consolidated-drained triaxial test of samples taken from the post-pampeano formation and the constitutive model calibration. Soft soil creep has a limitation to reproduce the mechanical behavior in low pressures in low strains. Nonetheless in this particular problem, the numerical model works in large strains and the proposed constitutive model reproduce reasonably the maximum deviatoric resistance of soft clays. The analyzed phenomena is not only dependent on the geotechnical parameters, but also has a strongly dependence with the riser-soil interaction. In this sense, it is important to define a proper friction angle for the interface. Several research papers has been published regarding this matter for driven pile, combining different materials such as concrete, wood and steel with different confining pressures. It has been proved that the friction angle for the steel-soil interaction is independent with the contact tension and the saturation degree and, for smooth steel, $\phi' = 24°$ (Potyondy 1961). This value is adopted for our numerical models.

Table 1. Constitutive models and parameters used.

| | Unit | Soft clay | Sandy soils |
|---|---|---|---|
| Model | - | Soft Soil Creep | HSsmall |
| Drainage | - | Undrained (A) | Drained |
| $\gamma$ | $kN/m^3$ | 16.0 | 19.0 |
| $\phi'$ | ° | 24 | 32 |
| $c'$ | $kPa$ | 0 | 0 |
| $\psi$ | ° | 0 | 0 |
| $C_c$ | - | 0.410 | - |
| $C_s$ | - | 0.080 | - |
| $C_\alpha$ | - | 0.013 | - |
| $G_0^{ref}$ | $MPa$ | - | 200 |
| $\gamma_{0.7}$ | - | - | $10^{-4}$ |
| $E_{ur}^{ref}$ | $MPa$ | - | 120 |
| $E_{50}^{ref}$ | $MPa$ | - | 40 |
| $E_{oed}^{ref}$ | $MPa$ | - | 40 |
| $m$ | - | - | 0.5 |
| $\nu_{ur}$ | - | 0.20 | 0.20 |
| OCR | - | 1.00 | 1.00 |
| $K_0^{nc}$ | - | 0.51 | 0.60 |
| $R_{inter}$ | - | 0.80 | 0.70 |
| $k$ | $10^{-6}$ m/s | 0.005 | 1.00 |

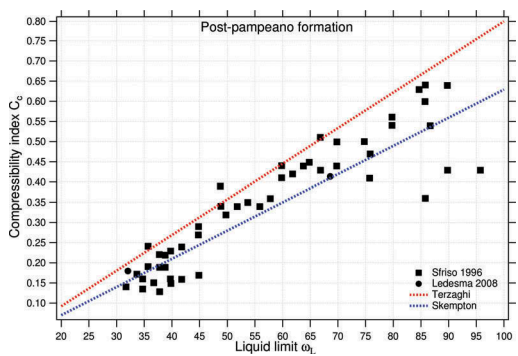

Figure 5. Compressibility index $C_c$ versus liquid limit for the post-pampeano formation. Comparisons with Terzaghi and Skempton expressions.

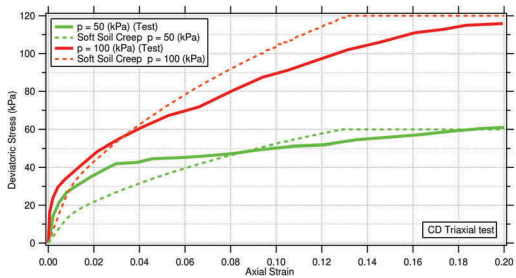

Figure 6. Constitutive model calibration and comparisons with CD triaxial test.

## 4 NUMERICAL MODELLING OF DOWNDRAG

### 4.1 Soil stress state after pipe jacking

Some authors have proposed theoretical and experimental models to estimate the strain path in the surrounding of a penetration test. Baligh (1985) provides a framework to study driven piles, assessing the disturbance of soil and showing that the strain produced by a driving maneuver is comparable to a radial cavity expansion since 2 radii from the tip (see Figure. 7). Other researchers arrived to similar conclusions (Pestana et al. 2002, Chong 2013). An axial-symmetric model is considered for our analysis, where a displacement equals to 35.5 cm, the riser radius, is imposed from the rotation axis. After this, a plate to model the metal sheet of the riser is used to simulate the soil-structure interaction. Figure 8 shows a detail of the mentioned stages. Table 2 details characteristics of each analyzed riser and the closest CPTu test, taken as reference to validate the geotechnical profile. It also summarize the sandy layers thickness crossed and riser length in both considered crown elevation. This data is used latter to correlate both, displacement and structural forces, with the presence of sands.

The assessed geotechnical profiles in each riser is plotted in Figure 9 for the former elevation, where soft clays are represented with green and sandy soils are represented with pink. Remolded clays, with a noticeable creep effect, are represented with grey and it occupies 2 diameter in riser's surrounding.

After the radial cavity expansion, the riser is installed and an excess pore pressure is generated in soft clays. Figure 10 shows the total pore pressure field in the former crown elevation, that has to be dissipated in the consolidation procedure. It is important to note that models for the actual crown elevation considers the same geotechnical profile, but ends at 28.0 m depth.

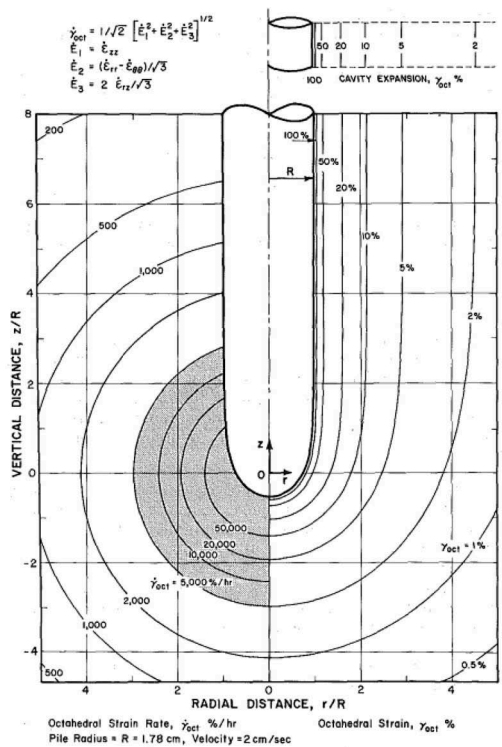

Figure 7. Total Strain path in a driven pile surroundings and comparison with a radial cavity expansion (Baligh, 1985).

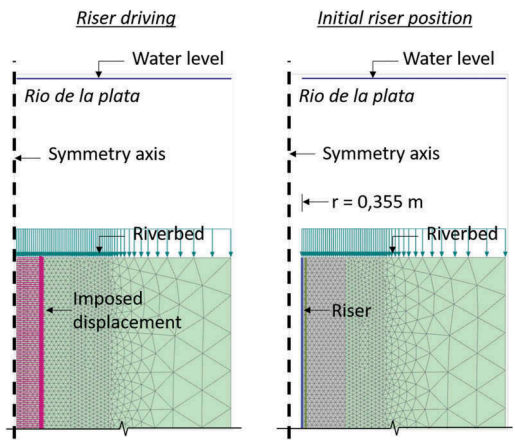

Figure 8. Construction stages of the finite element model.

### 4.2 Comparisons of numerical results with CPTu test

Obtained total pore pressures in our numerical models are compared with CPTu reference tests performed closest to the their future place, and the results are presented in Figure 11. Informed pore pressures are surveyed in the soil-riser interface.

Table 2. Characteristics of analyzed risers and reference undrained Cone Penetration Tests (CPTu).

| Riser N° | CPTu | New elevation | | Former elevation | |
|---|---|---|---|---|---|
| | | Sand layers thickness [m] | Riser length [m] | Sand layers thickness [m] | Riser length [m] |
| 1 | 01 | 2.7 | 28.0 | 5.3 | 30.5 |
| 3 | 01A | 1.0 | 28.0 | 1.0 | 30.5 |
| 5 | 01B | 2.7 | 28.0 | 5.2 | 30.5 |
| 9 | 02 | 3.5 | 28.0 | 6.0 | 30.5 |
| 12 | 03 | 2.8 | 28.0 | 5.3 | 30.5 |
| 16 | 04 | 1.9 | 28.0 | 4.4 | 30.5 |
| 21 | 05 | 6.0 | 28.0 | 8.5 | 30.5 |
| 23 | 05A | 5.0 | 28.0 | 7.5 | 30.5 |
| 27 | 06 | 4.2 | 28.0 | 6.7 | 30.5 |
| 34 | 07 | 1.3 | 28.0 | 3.8 | 30.5 |

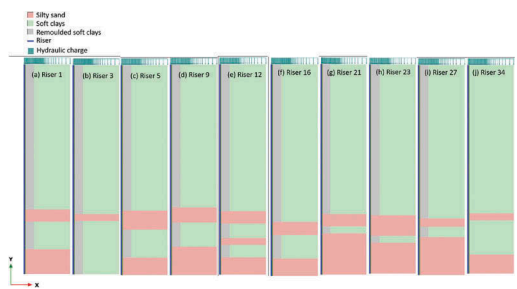

Figure 9. Geotechnical profiles proposed for each considered risers. Former crown elevation.

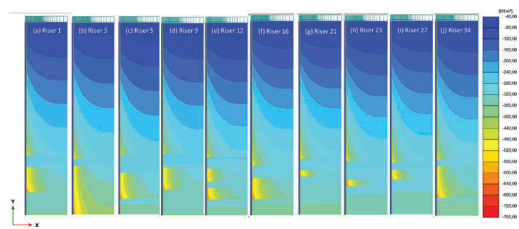

Figure 10. Numerical model results of total pore pressure in each riser. Former crown elevation.

It is interesting that our proposed model, despite its simplicity, reproduce accurately the excess pore pressure measured with CPTu. Whether the cone penetration test is performed or the riser is driving, produce a relatively similar pore pressure profile due to the strain range at both cases are working. As is shown in Figure 6, after 12% of axial strain, the deviatoric stress reaches a plateau and, for an undrained test, the pore pressure is nearly stabilized.

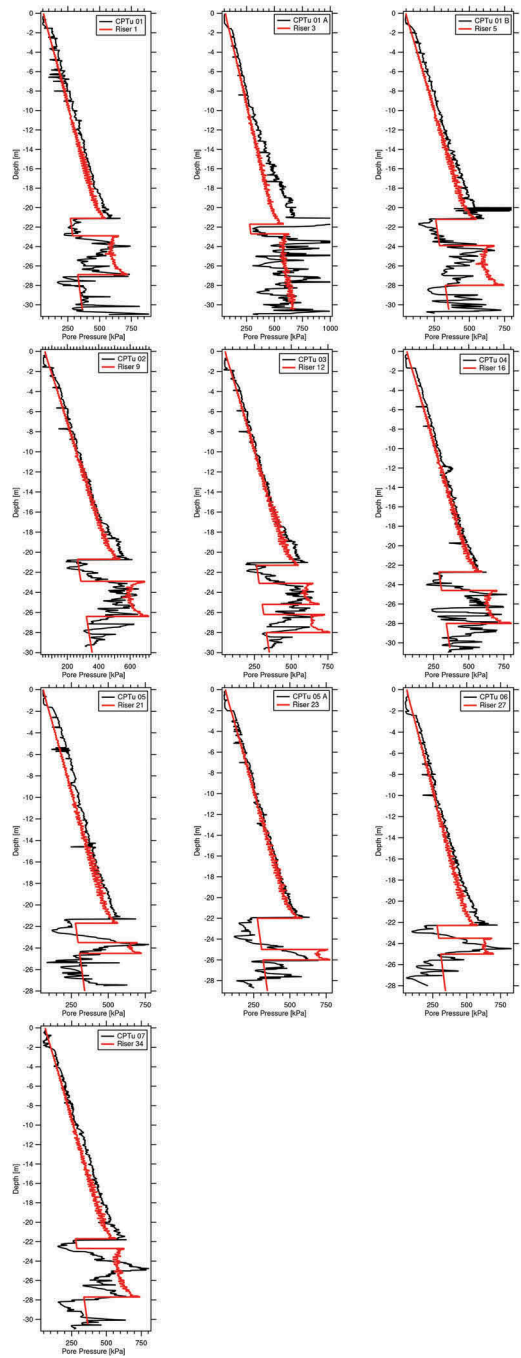

Figure 11. Comparison between CPTu test (black line) and numerical results obtained after a radial cavity expansion (red line).

After riser collocation, a consolidation stage is calculated considering a lifetime of 100 years. As was described in the introduction, two scenarios are studied in the risk analysis: a non bolted riser is analyzed in Scenario N°1, allowing a vertical degree of freedom in the pipe to be displaced inside the tunnel. This case is presented in Section 4.3. Scenario N°2, presented in Section 4.4, considers a vertical fixity in riser base to calculate the applied force in the launching ring.

### 4.3 Scenario 1 - Time evolution of riser downdrag

Riser vertical displacement is conceptually given by

$$\delta = \delta_e + \delta_c + \delta_{cr}, \tag{1}$$

where $\delta$ is the total vertical displacement, $\delta_e$ is the elastic bouncing right after the driving maneuver, $\delta_c$ is the consolidation component and $\delta_{cr}$ is the creep component. It is well known that elastic bouncing displacement in driving piles is negligible. In this way, the $\delta_e$ component is not considered in our paper focusing our curiosity in time-dependent terms.

Riser displacements over time due to primary consolidation $\delta_c$ (solid line) and due to primary and secondary $\delta_c + \delta_{cr}$ (dashed line), are plotted in Figure 12 for each considered riser in former crown elevation. Results obtained for the same analysis but in actual crown elevation, are presented in Figure 13. A hypothetical geotechnical profile where the riser does not cross any sandy layer, named 'pure soft clay', is included in both cases to estimate an upper bound for the results.

When sand lenses are coarser, displacement are less than 3 mm after 100 years in all cases with the exception of riser N°3, where there are not sands over the tunnel crown, neither in actual or former elevation. In this case, displacements in the end of the lifetime is 35 mm and the 90% of the value is reached after 3 month since its installation. A completely different setting is observed for the actual crown elevation, where risers N°1, 3, 5, 9, 12, 16 and 34 have maximum displacements higher than 10 mm, being a critical condition for the local hydraulic loss. Similarly to riser N°3 in former elevation, 90% of maximum displacements are reached after 2 or 3 month since their installation. Collecting all maximum displacement obtained in both elevations and ordering them in terms of the sandy soil thickness crossed, it can be seen in Figure 14 that, while sands thickness increases over 4.0 m, values are stabilized between 1 and 2 mm and the creep effect is almost negligible because the confinement effect produced by coarser soils. The mentioned effect produced by sands generates an increment in structural forces in pipes. Figure 15 shows the maximum normal force reached in risers in terms of the sand soil thickness crossed, for both elevation. It is clear that, when the displacement freedom is blocked by some reason, the structural element is more stressed. When sand layers are less than 3 m, the maximum normal force is almost constant between 900 and 1400 kN. From 3 m and up, the normal compression increase dramatically, reaching values nearly to 2800 kN. Different from the displacements, the presence of sandy soils has not a blocking effect of creep in the structural forces.

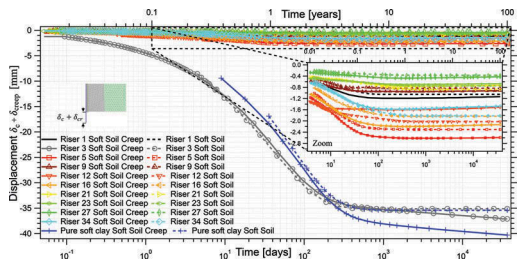

Figure 12. Riser displacement due to primary and secondary consolidation in former crown elevation.

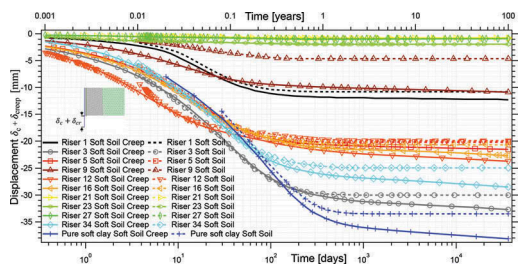

Figure 13. Riser displacement due to primary and secondary consolidation in actual crown elevation.

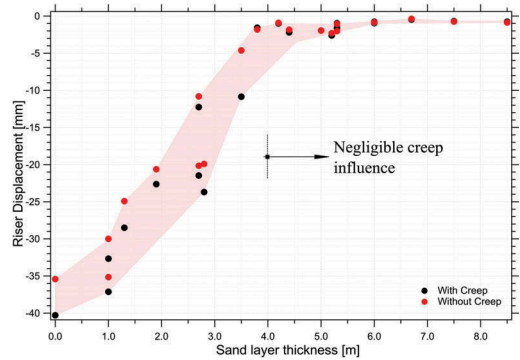

Figure 14. Maximum displacements in terms of sand soil thickness crossed by riser in both elevations in the end of the lifetime.

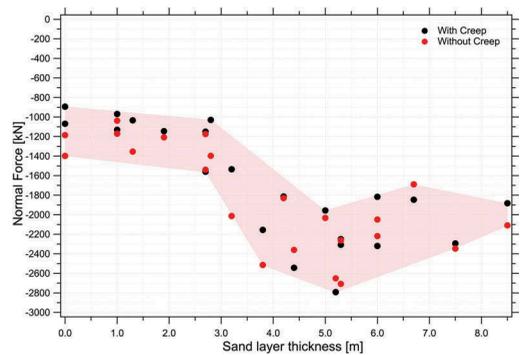

Figure 15. Maximum normal force in riser in terms of sand soil thickness crossed by riser in both elevations in the end of the lifetime.

### 4.4 Scenario 2 - Structural forces induced by consolidation

When risers are bolted to the launching ring, the vertical displacement is restricted and consolidation and creep of soft clays generates a concentrated force into the crown lining segment. This new scenario produce a rearrangement of internal stresses in pipes.

Figure 16 shows maximum normal forces for riser-tunnel union (shaded in blue) and for all pipes (shaded in red), in the end of the lifetime, in terms of crossed sand layer thickness. When sand thickness is less than 3 m, the maximum normal force is placed in the riser-tunnel union and are almost three time higher than the maximum normal force reached in scenario N°1. While the sand thickness increases, the force applied in the riser-tunnel union decreases to the half or less, but the maximum normal force in all over the riser is remains almost unalterable. In order to illustrate the internal distribution of normal and circumferential structural forces of two extrt4eme cases, diagrams for risers N°1 and N°3 are plotted in Figures 17 and 18 respectively. It can be seen in all cases that normal forces in the first day from the installation are related with its own weight and, while time elapses, normal compression increases considerably. In circumferential normal diagrams the opposite effect occurs, being higher right after the riser driving and decreasing with time.

When presence of sandy soils are considerably, normal forces in both scenarios are almost equals, increasing in scenario N°2 on the vicinity of the tunnel, being the maximum value is within the transition layer. If the geotechnical profile is composed mainly with soft clays, differences between both scenarios are notorious and the maximum normal value is in the tunnel crown. Circumferential forces has their maximum values coincidentally with sand layers and differences between scenarios are negligible no matter the presence of sand layers, but being clear its contribution to the riser confinement.

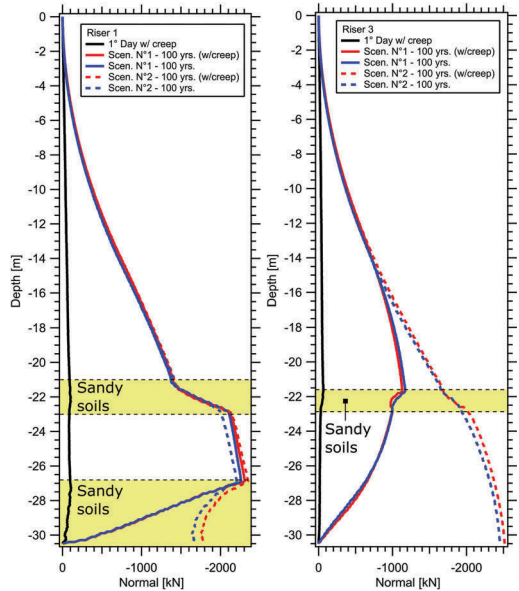

Figure 17. Normal forces for risers N°1 and N°3 in both scenarios.

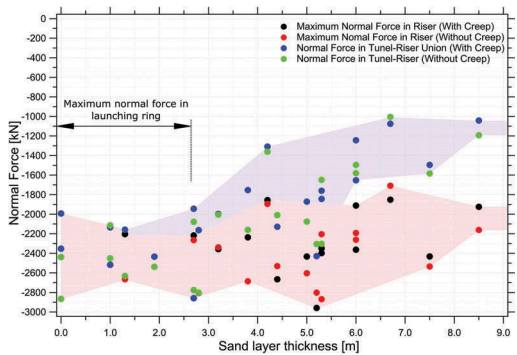

Figure 16. Maximum normal forces in risers in terms of sand soil thickness crossed in both elevations in the end of the lifetime.

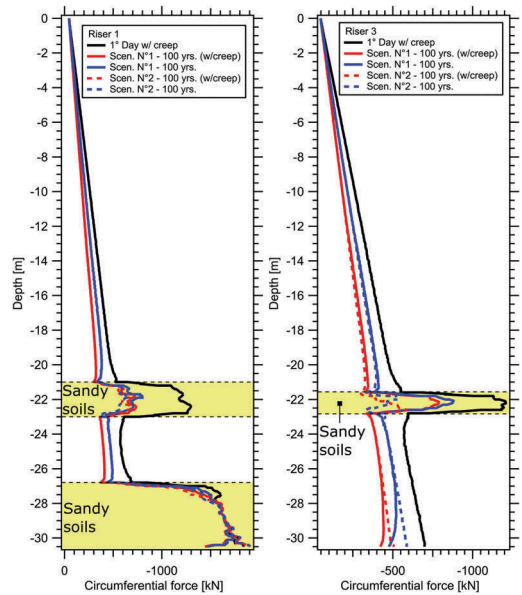

Figure 18. Circumferential forces for risers N°1 and N°3 in both scenarios.

## 5 CONCLUSIONS

The driving procedure of piles and tubes it is still a challenging task to scientist and engineers. In this paper, a simple but powerful finite element model has been proposed to estimate the downdrag phenomena and the negative friction due to consolidation and creep, calibrating the soil profile using CPTu test. Our proposal deals with the identification and modeling of a layered stratum, characterizing the transition as a composition of simple geotechnical units and calibrated comparing the pore pressure obtained in our numerical models with those obtained in field tests, simulating the penetration strain path as a radial cavity expansion.

After the riser driving, a negative skin friction in the soil-structure interface due to the consolidation and creep is generated due to clay remolding, moving the pipe inside of the tunnel in Scenario N°1. In Scenario N°2, where the riser is bolted in the launching ring, the mentioned phenomenon produce a non-negligible concentrated force in the crown tunnel.

In both scenarios, displacements or the forces, has a strongly dependence with the sand layer crossed by the riser. In Scenario N°1 a maximum displacement of 3.5 cm is reached by Riser N°3, and it was proved that a lifting in the elevation of the tunnel implies an increasing movement in the rest of pipes, due to a less thickness in crossed sand layers. It was also showed that an increment in sand layers rebounds in a considerable reduction of the creep component and, after 60 to 90 days, the majority of the movement has been performed. When sand layer thickness is larger than 4 m, the riser displacement is negligible.

If the pipe is bolted to the launching ring, the force expected in the union range from 2900 kN to 1000 kN, being also the maximum force within the riser if the sand layer thickness crossed is less than 2.5 m. While the thickness increases, the force in the tunnel crown decrease and the maximum normal force is moved to the transition layer.